\documentstyle[12pt,amssymb]{article}

\def\fnote#1#2{\begingroup\def\thefootnote{#1}\footnote{#2}\addtocounter{footnote}{-1}\endgroup}

\def\inbar{\vrule height1.5ex width.4pt depth0pt}
\def\IB{\relax{\rm I\kern-.18em B}}
\def\IC{\relax\,\hbox{$\inbar\kern-.3em{\rm C}$}}
\def\ID{\relax{\rm I\kern-.18em D}}
\def\IE{\relax{\rm I\kern-.18em E}}
\def\IF{\relax{\rm I\kern-.18em F}}
\def\IG{\relax\,\hbox{$\inbar\kern-.3em{\rm G}$}}
\def\IH{\relax{\rm I\kern-.18em H}}
\def\II{\relax{\rm I\kern-.18em I}}
\def\IK{\relax{\rm I\kern-.18em K}}
\def\IL{\relax{\rm I\kern-.18em L}}
\def\IM{\relax{\rm I\kern-.18em M}}
\def\IN{\relax{\rm I\kern-.18em N}}
\def\IO{\relax\,\hbox{$\inbar\kern-.3em{\rm O}$}}
\def\IP{\relax{\rm I\kern-.18em P}}
\def\IQ{\relax\,\hbox{$\inbar\kern-.3em{\rm Q}$}}
\def\IR{\relax{\rm I\kern-.18em R}}
\def\IT{\relax{\rm I\kern-.18em T}}
\def\ZZ{\relax{\sf Z\kern-.4em Z}}

      \def\l{\lambda}

  \def\cI{{\cal I}} 
   
\def\cO{{\cal O}} \def\cP{{\cal P}}  \def\cR{{\cal R}}

\def\mathC{{\mathbb C}}  \def\mathF{{\mathbb F}}
  \def\mathP{{\mathbb P}} \def\mathQ{{\mathbb Q}}
  \def\mathZ{{\mathbb Z}}

\def\Ewhat{{\widehat E}}

\def\fnote#1#2{\begingroup\def\thefootnote{#1}\footnote{#2}\addtocounter
{footnote}{-1}\endgroup}
\def\beq{\begin{equation}}
\def\eeq{\end{equation}}
\def\bea{\begin{eqnarray}}
\def\eea{\end{eqnarray}}
\def\llea#1{\label{#1}\eea}
\def\lleq#1{\label{#1}\eeq}
\let\nn=\nonumber
\def\tabroom{\hbox to0pt{\phantom{\Huge A}\hss}}
\def\notin{\ \hbox{{$\in$}\kern-.51em\hbox{/}}}

  \def\E1Fq{E_1/\IF_q}

     \def\rmGSO{{\rm GSO}}  
\def\rmHW{{\rm HW}}       
     
\def\rmSL{{\rm SL}}      

\def\pwhat{{\widehat p}}

\def\Awhat{{\widehat A}}

\def\cOwhat{{\widehat \cO}}  \def\cIwhat{{\widehat \cI}}  \def\cRwhat{{\widehat \cR}}

\def\notdiv{{\relax{~|\kern-.35em /~}}}

\def\boxit#1{
\vbox{\hrule height1pt\hbox{\vrule width1pt\kern0.3cm
\vbox{\kern0.3cm\hbox{$\displaystyle#1$}\kern0.3cm}\kern0.3cm\vrule
width1pt}\hrule height1pt}}

\begin{document}
\parindent=0pt

\hfill \phantom{\today}

\vskip 0.9truein

 \centerline{{\bf A MODULARITY TEST FOR ELLIPTIC MIRROR SYMMETRY}}

\vskip .4truein

 \centerline{\sc Rolf Schimmrigk\fnote{\dagger}{email: netahu@yahoo.com}}

\vskip .3truein

\centerline{\it Indiana University South Bend}
 \vskip .05truein
\centerline{\it 1700 Mishawaka Ave., South Bend, IN 46634}

\vskip 1.2truein

\baselineskip=19pt

\centerline{\bf ABSTRACT:}

\vskip .2truein

\begin{quote}
 In this note a prediction of an algebraic mirror construction is
 checked for elliptic curves of Brieskorn-Pham type via number theoretic methods.
 It is shown that the modular forms associated to the Hasse-Weil L-series
 of mirror pairs of such curves are identical.
\end{quote}

\vfill

{\sc PACS Numbers and Keywords:} \hfill \break Math:  11G25
Varieties over finite fields; 11G40 L-functions; 14G10  Zeta
functions; 14G40 Arithmetic Varieties \hfill \break Phys: 11.25.-w
Fundamental strings; 11.25.Hf Conformal Field Theory; 11.25.Mj
Compactification

\renewcommand\thepage{}
\newpage


\baselineskip=21.4pt

\parskip=.15truein
\parindent=0pt
\pagenumbering{arabic}

\section{Introduction}

Mirror symmetry is usually considered in the context of varieties
defined over complete fields, such as the complex number field
$\mathC$. It has become clear in the recent past that interesting
information about the underlying string physics can be obtained by
considering manifolds over fields of different type, in particular
those of positive characteristics, because this allows to probe
the geometry with tools that are not available over the complex
numbers. It turns out, in particular, that certain generating
functions associated to such arithmetic probes are directly
related to generating functions on the string worldsheet, thereby
connecting the physics on the string worldsheet theory to
arithmetic quantities derived from the geometry of the
corresponding varieties. More precisely, it was shown in refs.
\cite{su03, ls04, s05, s06} that it is possible to relate modular
forms constructed from the worldsheet conformal field theory to
modular forms that arise from the arithmetic of the associated
varieties. It was furthermore shown that it is possible to
reconstruct Brieskorn-Pham type Calabi-Yau manifolds directly from
the modular forms derived from conformal field theories \cite{s05,
s06}.

The main motivation for the program developed in the above
references has been to gain a better understanding of the
emergence of spacetime in string theory. This is an old problem
that has previously resisted a concrete formulation amenable to
explicit constructions. A second motivation comes from the hope
that a more incisive understanding of the relation between the
geometry of spacetime and the physics of the worldsheet might also
lead to a better understanding of mirror symmetry. It is in the
context of exact models that mirror symmetry is represented by
very simple operations, and the question becomes whether these
operations can be mapped into the geometric framework. The problem
of an interpretation of mirror symmetry in the context of the
congruence zeta function of Artin has been explored in refs.
\cite{cdov00, cdov03}, and has been further discussed in
\cite{y01,k04,k06, ky06}. This has turned out to be a difficult
problem because cycles of different dimensions are encoded in the
Artin zeta function in rather different ways. It is therefore
natural to ask whether the construction of worldsheet quantities
using the methods considered in refs. \cite{su03, ls04, s05, s06}
can be used to probe mirror symmetry in a different way. The
purpose of this note is to use the physical interpretation of the
Hasse-Weil L-function found in these papers to provide a string
theoretic modularity test of the geometric mirror construction
described in refs. \cite{ls90b, ls97}.

The outline of this article is as follows. Section 2 briefly
describes the ring isomorphism introduced in \cite{ls90b, ls97}.
In Section 3 this isomorphism is applied to the construction of
the algebraic mirrors of the elliptic cubic Fermat curve $E^3
\subset \mathP_2$. The mirror curves are described by polynomials
that are not diagonal, and live in weighted projective spaces. A
priori their Hasse-Weil L-functions thus should be expected to be
different. The conformal field theory $C^3$ related to the curve
$E^3$ is, however, isomorphic to the theory corresponding to the
mirror curve. The worldsheet modular forms that were found in
\cite{su03} to provide the building blocks for the elliptic
modular form of the Fermat cubic should therefore also enter the
modular form of the mirror. Hence the L-function of the algebraic
mirror should be identical to that of $E^3$. This is confirmed in
Section 4. A similar analysis can be applied to the remaining
elliptic curves of Brieskorn-Pham type, using the results of
\cite{ls04, s05}.

\section{Geometric mirror map}

The following paragraphs contain a brief review of the algebraic
mirror map introduced in \cite{ls90b,ls97} to make this article
self-contained. The original motivation for this map arose from
the observation of cohomological mirror symmetry in the context of
weighted projective hypersurfaces described in \cite{cls90}. In
that construction, or rather in the Landau-Ginzburg version of
this class, the computation of the Hodge numbers produced a highly
symmetric distribution of these numbers, and the question arose
whether "Hodge mirror pairs" are actually mirrors. In the class of
varieties considered in \cite{cls90} there appeared e.g. two
spaces with Hodge numbers that are mirrors of the Hodge numbers of
the quintic. The question then was whether this was accidental or
based on an isomorphism of the associated conformal field theory.
The strategy of refs. \cite{ls90b, ls97} to address this issue was
to establish an isomorphism between two differently constructed
orbifold rings, which emerge as building blocks in different
Calabi-Yau varieties.
 The original idea for this construction arose from the
 observation in \cite{ls90a} that the transition from diagonal
 affine invariants to D-type invariants in the partition function
 of Gepner models sometimes produces mirror theories.
The basic isomorphism of \cite{ls90b, ls97} is a generalization of
this transition to much more general Landau-Ginzburg theories.

The first ring is defined via the quotient
 \beq
  \cR = \mathC_{\left(\frac{b}{g_{ab}}, \frac{a}{g_{ab}}\right)}
   [x_1,x_2]/\cI_{a,b},
 \eeq
 where the ideal $\cI_{a,b}$ is defined by the polynomial
  \beq
  p_{a,b}(x_1,x_2) = x_1^a + x_2^b.
  \eeq
 Here $g_{ab}$ is the greatest common divisor of $a$ and $b$.
 The orbifold ring $\cO$ derived from $\cR$ is then defined
 via the cyclic group $G_b = \mathZ/b\mathZ$ as
 \beq
  \cO = \cR/G_b,
 \eeq
 where the action of $G_b$ is defined as
 \beq
 A: ~(x_1,x_2) \mapsto (\xi^{b-1}x_1, \xi x_2).
 \lleq{A-action}

The second ring is constructed via
 \beq
 \cRwhat = \mathC_{\left(\frac{b^2}{h_{ab}},\frac{a(b-1)-b}{h_{ab}}
 \right)}[y_1,y_2]/\cIwhat_{ab},
 \eeq
 where the ideal $\cIwhat_{a,b}$ is defined by the polynomial
 \beq
 \pwhat_{a,b}(y_1,y_2) = y_1^{a(b-1)/b} + y_1y_2^b,
 \eeq
 and $h_{ab}$ denotes the greatest common divisor of $b^2$
 and $(ab-a-b)$. The orbifold ring $\cOwhat$ derived from $\cRwhat$ is
 defined as
 \beq
 \cOwhat = \cRwhat/G_{b-1},
 \eeq
 where the action of $G_{b-1}$ is defined as
 \beq
 \Awhat:~(y_1,y_2) \mapsto (\xi y_1, \xi^{b-2}y_2).
 \eeq

One can show that there exists a 1-1 transformation that maps
these orbifold rings into each other \cite{ls97}, leading to the
following result.

{\bf Proposition.} {\it The rings $\cO$ and $\cOwhat$ are
isomorphic.}

More explicitly, the basic isomorphism can be summarized as
 \bea
  & &\IC_{\left(\frac{b}{g_{ab}},\frac{a}{g_{ab}}\right)}
                                            \left[\frac{ab}{g_{ab}}\right]
    \ni \left \{z_1^a+z_2^b=0\right \}
                 ~{\Big /}~ G_b: \left[\matrix{(b-1)&1}\right] ~~
                                              \nn \\ [3ex]
&\cong &
\IC_{\left(\frac{b^2}{h_{ab}},\frac{a(b-1)-b}{h_{ab}}\right)}
                                         \left[\frac{ab(b-1)}{h_{ab}}\right]
       \ni \left \{y_1^{a(b-1)/b}+y_1y_2^b=0\right\}
                 ~{\Big /}~ G_{b-1}:
                 \left[\matrix{1&(b-2)}\right],
 \llea{basic-iso}
 where $G_b: [(b-1)~1]$ denotes the group element of $G_b$ acting
 as defined in (\ref{A-action}).

This basic isomorphism can be applied to algebraic varieties of
any dimension, where it can lead to a number of different
phenomena, depending on how the quotient construction involved
combines with the (weighted) projective invariance of the ambient
space. It can happen, in particular, that the symmetries of the
affine surface defining the quotients of the image theory become
part of the weighted projective equivalence when the singularities
just described are embedded in Calabi-Yau varieties. The resulting
spaces can then become mirrors of each other if the resolution of
the singularities produces the appropriate cohomological
structure.

The simplest application of the strategy just outlined is provided
by  3-folds for which the basic isomorphism itself gives the
mirror map, without the necessity of an iterative application.
Such an example is provided by the mirror configuration
 \beq
 \mathP_{(3,8,33,66,88,132)}[264]^{(57,81)}/G_2
   \sim \mathP_{(3,8,66,88,99)}[264]^{(81,57)},
   \eeq
 discussed in \cite{ls90a}.

In general an iterative application of the basic map is necessary
to construct the mirror manifold. This can be illustrated by
constructing the algebraic mirror of the quintic threefold family
 \beq
 X_{\l} = \left\{(z_0:\cdots :z_4)\in \mathP_4~{\Big |}~
 \sum_i z_i^5 + \l \prod_i z_i =0 \right\},
 \eeq
 by applying the ring isomorphisms iteratively as follows. The
 quotient construction of the exact
model mirror \cite{gp90} suggests to consider the quotient by the
product of four cyclic groups $G_5$ via the following action
 \beq
 G_5^4:~~\left[\matrix{4 &1 &0 &0 &0\cr
                        0 &4 &1 &0 &0\cr
                        0 &0 &4 &1 &0\cr
                        0 &0 &0 &4 &1\cr}\right]
 \lleq{quintact}
where the notation $[0~0~4~1~0]$ e.g. means the action
 \beq
(z_0,z_1,z_2,z_3,z_4) \mapsto (z_0,z_1,\xi^4z_3,\xi z_3,z_4),
\eeq
 where $\xi \in \mu_5$ is a fifth root of unity. The ring
isomorphisms (\ref{basic-iso}) then lead to the mirror manifold
 \beq
 X_{\l}' = \left\{(y_0:\cdots :y_4)\in \IP_{(64,48,52,51,41)}~|~ y_0^4+
y_0y_1^4+y_1y_2^4 + y_2y_3^4+y_3y_4^5 + \l \prod_i y_i =0
\right\}.
 \eeq
 These examples generalize to many families in the class of
 weighted projective space constructed in \cite{cls90, ks92, krsk92}.

\section{Mirror families of deformed elliptic Brieskorn-Pham curves}

 Denote by $A^{(1)}_{1,k}$ the affine Lie algebra at conformal
 level $k$ based on $\rm sl(2,\mathC)$, and let $C^d$ for $d=3,4,6$ be
 the three exactly solvable conformal field theory models at central
 charge $c=3$. They are given by the GSO projected tensor products of
 $N=2$ minimal superconformal models based on $A^{(1)}_{1,k}$
  \cite{g88}
 \bea
  C^3 &=& (A^{(1)}_{1,1})^{\otimes 3}_{\rmGSO} \nn \\
  C^4&=& (A^{(1)}_{1,2})^{\otimes 2}_{\rmGSO} \nn \\
  C^6 &=& (A^{(1)}_{1,1} \otimes A^{(1)}_{1,4})_{\rmGSO},
 \eea
 The notation of these
 models is motivated by the Landau-Ginzburg analyses of refs. \cite{m89,
 vw89, lvw89}, according to which these models are expected to
 be related to the elliptic Brieskorn-Pham curves
 \bea
 E^3 &=& \left\{(z_0:z_1:z_2) \in \IP_2~|~z_0^3+z_1^3+z_2^3=0 \right\}
                        \nn \\
 E^4 &=& \left\{(z_0:z_1:z_2) \in \IP_{(1,1,2)}~|~z_0^4+z_1^4+z_2^2=0 \right\}
                 \nn \\
 E^6 &=& \left\{(z_0:z_1:z_2) \in \IP_{(1,2,3)}~|~z_0^6+z_1^3+z_2^2=0 \right\}
 \llea{elliptic-bp-curves}
It was shown in \cite{s05} that the curves $E^d$ can be
constructed directly from the conformal field theories $C^d$ via
the modular forms that enter their partition functions.

Applying the isomorphism (\ref{basic-iso}) to the cubic family
 \beq
 E^3_{\psi} = \left\{(z_0:z_1:z_2) \in \mathP_2~|~
  z_0^3 + z_1^3 + z_2^3 - 3\psi z_0z_1z_2 = 0 \right\}
  \eeq
leads to the elliptic mirror family $\Ewhat^3_{\psi} =
E^3_{\psi}/G_3$, where the action of the group $G_3\cong \mu_3$ is
given by
 \beq
  G_3: (z_0,z_1,z_2) \mapsto  (\xi_3^2z_0, \xi_3z_1, z_2).
 \eeq
  The group action of the image theory under the basic isomorphism
  becomes part of the weighted projective equivalence, hence
  the algebraic image of this mirror is given by
 \beq
 \Ewhat^3_{\psi} = \left\{(z_0:z_1:z_2) \in \mathP_{(3,1,2)}~|~
  z_0^2 + z_0z_1^3 + z_2^3 - 3\psi z_0z_1z_2 = 0 \right\}.
  \eeq
  There exists a second quotient, which can be constructed by
  considering two $\mu_3$ groups, with an action given by
 \beq
 G_3^2:~~\left[\matrix{2&1&0\cr
                         0&2&1\cr}\right].
 \eeq
 Applying the basic ring isomorphism
iteratively leads to the algebraic form of the mirror given by
\beq
 \Ewhat^{3'}_{\psi} = \left\{(z_0:z_1:z_2)\in \mathP_{(2,1,1)}~|~ z_0^2 +
z_0z_1^2+z_1z_2^3 - 3\psi z_0z_1z_2 = 0\right\}.
 \eeq

The results of \cite{s05} show that the L-functions of the
Brieskorn-Pham curves $E^6 \subset \mathP_{(1,2,3)}$ and $E^4
\subset \mathP_{(1,1,2)}$ are different from the L-function of
$E^3$. Because $\Ewhat^3$ and $\Ewhat^{3'}$ are algebraic mirrors
of the Fermat cubic $E^3$, their L-functions should agree with the
L-function of $E^3$, even though they are hypersurfaces in the
same weighted projective planes as $E^6$ and $E^4$, respectively.

Similar considerations apply to the algebraic image of the
quotient of the quartic family $E^4_{\psi}$ and the degree six
family $E^6_{\psi}$. For $E^4_{\psi}$ the algebraic mirror map
leads to the family
 \beq
 \Ewhat^4_{\psi} =
\{(z_0:z_1:z_2)\in \mathP_{(2,1,3)}~|~ z_0^3 + z_0z_1^4 + z_2^2
 - 4\psi z_0z_1z_2 = 0\},
 \eeq
 while the algebraic mirror of $E^6_{\psi}$ is
 \beq
 \Ewhat^6_{\psi} = \{(z_0:z_1:z_2)\in \mathP_{(1,1,2)}~|~
   z_0^4+z_0z_1^3 + z_2^2  - 6\psi z_0z_1z_2 = 0\}.
 \eeq
 In both cases the symmetry actions on the image theories are
 trivial again.

For these examples mirror symmetry again predicts that the
L-functions of $\Ewhat^4 \subset \mathP_{(1,2,3)}$ and $\Ewhat^6
\subset \mathP_{(1,1,2)}$ should agree with the L-functions of
$E^4 \subset \mathP_{(1,1,2)}$ and $E^6 \subset \mathP_{(1,2,3)}$
respectively, even though the L-functions of the Brieskorn-Pham
points in these weighted ambient spaces are different. The
expectations from the algebraic mirror map are confirmed in the
next section.

\section{The Hasse-Weil L-Function of Elliptic Mirror Pairs}

A detailed review to the Hasse-Weil L-function can be found in
many references, e.g. \cite{l96} (a summary with focus on elliptic
curves can be found in \cite{s05}). Briefly, the Hasse-Weil
L-function of an algebraic curve $X$ is determined by the local
congruence zeta functions at all prime numbers $p$. This is
defined in \cite{w49} as the generating series
 \beq
 Z(X/\mathF_p, t) =
 \exp\left(\sum_{r\in \IN} N_{r,p}(X) \frac{t^r}{r}\right),
 \eeq
 where $N_{r,p}(X)=\# \left(X/\mathF_{p^r}\right)$ denotes the
 cardinality of the variety over the finite extension
 $\mathF_{p^r}$ of the finite field $\mathF_p$ of characteristic
 $p$ for any rational prime $p$, and $t$ is a formal variable.
 It is a classic result by F.K. Schmidt that $Z(X/\mathF_p,t)$ is a
 rational function, taking the form
 \beq
 Z(X/\mathF_p,t) = \frac{\cP_p(t)}{(1-t)(1-pt)},
 \eeq
 where $\cP_p(t)$ is a quadratic polynomial.

 More relevant from a physical perspective than
the local zeta functions is the global zeta function, obtained by
setting $t=p^{-s}$ and taking the product over all rational primes
at which the variety has good reduction. Denote by $S$ the set of
rational primes at which $X$ becomes singular and denote by $P_S$
the set of primes that are not in $S$. The global zeta function
can be defined as
 \beq
  Z(X,s) = \prod_{p \in P_S}
\frac{\cP_p(p^{-s})}{(1-p^{-s})(1-p^{1-s})} =
\frac{\zeta(s)\zeta(s-1)}{L(X,s)},
 \lleq{globalzeta}
  where the Hasse-Weil L-function has been introduced as
 \beq
  L(X,s) \doteq \prod_{p \in P_S} \frac{1}{\cP_p(p^{-s})},
  \eeq
   and
 $\zeta(s) = \prod_p (1-p^{-s})^{-1}$ is the Riemann zeta function
 of the
rational field $\mathQ$. Here $\doteq$ denotes the L-function up
to a finite number of primes. The treatment of the finite number
of exceptional primes is more elaborate and can be found e.g. in
\cite{s05}, leading to the completion of the L-functions of the
curves $E^d$ at those primes for which they are singular.

In refs. \cite{su03,ls04,s05} the Hasse-Weil L-functions
$L_{\rmHW}(E^d,s)$  of the curves $E^d, d=3,4,6$ were computed
explicitly, leading to three distinct series. It was shown that
all three corresponding modular cusp forms of weight two factor
into string theoretic forms twisted by characters that are
associated to number fields that are determined by the quantum
dimensions of the conformal field theory. More precisely, denote
by $S_2(\Gamma_0(N))$ the space of cusp forms of modular level $N$
and weight two with respect to the congruence group $\Gamma_0(N)
\subset \rmSL(2,\mathZ)$. Then the modular forms associated to the
L-series of $E^d$ are elements $f(E^d,q) \in S_2(\Gamma_0(N_d))$
with $N_d=27,64,144$ for $d=3,4,6$ respectively. Their string
theoretic factorizations take the form
  \bea
 f(E^3,q) &=& \Theta^1_{1,1}(q^3)\Theta^1_{1,1}(q^9) \nn \\
 f(E^4,q) &=& \Theta^2_{1,1}(q^4)^2\otimes \chi_2 \nn \\
 f(E^6,q) &=& \Theta^1_{1,1}(q^6)^2\otimes \chi_3
 \llea{hw-as-thetas}
 where
  \beq
  \Theta^k_{\ell,m}(\tau)=\eta^3(\tau)c^k_{\ell,m}(\tau)
  \eeq
   are
 Hecke indefinite modular forms constructed from Kac-Peterson
 string functions $c^k_{\ell,m}(\tau)$ associated to the affine Lie algebra
 $A^{(1)}_1$ \cite{kp84}. This shows that the arithmetic method is
precise enough to detect the different details of the underlying
conformal field theory, even in the elliptic framework.

The factors $\cP_p(t)$ can be obtained by expanding Weil's
defining form of the congruence zeta function and comparing the
coefficients to the expansion of Schmidt's rational form of it.
Writing the polynomials $\cP_p(t)$ at the good primes as
 \beq
  \cP_p(t)= 1 + \beta_1(p)t + pt^2,
 \eeq
 the coefficient $\beta_1(p)$ is
 expressed in terms of the cardinalities $N_{1,p}= \#(X/\mathF_p)$ as
 \beq
 \beta_1(p) = N_{1,p} - (p+1).
  \eeq
 The elliptic modularity theory proven in \cite{w95, bcdt01}
 guarantees that the computation of a finite number of cardinalities
 is sufficient to determine the corresponding modular
forms completely. For the elliptic mirror curve $\Ewhat^3$ the
results for the numbers $N_{1,p}$ for the first few primes are
collected in Table 1.
\begin{center}
\begin{tabular}{l| r r r r r r r r r r r}
Prime $p$    &2 &3 &5  &7  &11  &13    &17  &19  &23  &29  &31
\tabroom \\
\hline $N_{1,p}$    &3 &4 &6  &9  &12  &9     &18  &27  &24  &30
&36
\tabroom \\
\hline $\beta_1(p)$ &0 &0 &0  &1  &0   &$-5$  &0   &7   &0   &0
&4
\tabroom \\
\hline
\end{tabular}
\end{center}

{\bf Table 1.}{\it ~~The coefficients
$\beta_1(p)=N_{1,p}(\Ewhat^3)-(p+1)$ of the elliptic cubic curve
$\Ewhat^3$ in terms of the cardinalities $N_{1,p}$ for the lower
rational primes.}

Inserting these cardinalities into the Hasse-Weil series of the
mirror curve $\Ewhat^3$ of the cubic Fermat curve leads to
 \beq
 L_{\rmHW}(\Ewhat^3,s)
  = 1 - \frac{2}{4^s} - \frac{1}{7^s} + \frac{5}{13^s} +
       \frac{4}{16^s} -\frac{7}{19^s} + \cdots
 \eeq
Comparing this result with the Hasse-Weil L-function computed in
\cite{su03,s05} shows agreement. It can similarly be shown that
the L-function of the curve $\Ewhat^{3'}$ agrees with that of
$E^3$.

 The mirrors $\Ewhat^4$ and $\Ewhat^6$ furthermore have the same L-function
 as $E^4$ and $E^6$, respectively. Summarizing, the curves $E^d$ have the
 property that  $L_{\rmHW}(\Ewhat^d,s) = L_{\rmHW}(E^d,s)$.
 Expressed in terms of the associated modular forms therefore
 leads to the following result.

 {\bf Proposition.} ~{\it The modular forms associated to
  mirror pairs $(E^d,\Ewhat^d)$ of elliptic Brieskorn-Pham
  curves $E^d$ of degree $d=3,4,6$ satisfy the identity}
  \beq
   f(\Ewhat^d,q) = f(E^d,q).
  \eeq

\vskip .3truein

{\bf ACKNOWLEDGEMENT.}

It is a pleasure to thank Monika Lynker for conversations and Jack
Morse for correspondence. Part of this work was supported by a KSU
Incentive Grant for Scholarship, and an IUSB Faculty Research
Grant.

\end{document}